\documentclass[11pt,twoside]{article}
\usepackage[pdftex]{graphicx}
\usepackage{amsmath}
\usepackage{amssymb}
\usepackage{cite}

\usepackage{mathrsfs}

\begin{document}
\newcommand{\pst}{\hspace*{1.5em}}

\newcommand{\rigmark}{\em Journal of Russian Laser Research}
\newcommand{\lemark}{\em Volume 30, Number 5, 2009}

\newcommand{\be}{\begin{equation}}
\newcommand{\ee}{\end{equation}}
\newcommand{\bm}{\boldmath}
\newcommand{\ds}{\displaystyle}
\newcommand{\bea}{\begin{eqnarray}}
\newcommand{\eea}{\end{eqnarray}}
\newcommand{\ba}{\begin{array}}
\newcommand{\ea}{\end{array}}
\newcommand{\arcsinh}{\mathop{\rm arcsinh}\nolimits}
\newcommand{\arctanh}{\mathop{\rm arctanh}\nolimits}
\newcommand{\bc}{\begin{center}}
\newcommand{\ec}{\end{center}}

\thispagestyle{plain}

\label{sh}


\begin{center} {\Large \bf
BOUND ENTANGLED STATES OF FOUR QUBITS IN TOMOGRAPHIC PROBABILITY REPRESENTATION
 } \end{center}

\bigskip

\bigskip

\begin{center} {\bf
{\bf V. I. Man'ko$^{*}$, I. V. Traskunov$^{**}$}
}\end{center}

\medskip

\begin{center}
{\it
$^{*}$P.~N.~Lebedev Physical Institute, Russian Academy of Sciences\\
Leninskii Prospect 53, Moscow 119991, Russia
}
\smallskip

{\it
$^{**}$Moscow Institute of Physics and Technology (State University)\\
Institutskii per. 9, Dolgoprudnyi, Moscow Region 141700, Russia\\
\smallskip
}
e-mail:~~~igor-michigan~@~yandex.ru
\end{center}

\begin{abstract}\noindent
The entanglement phenomenon on example of Smolin state of four qubits is discussed. This state is known as bound entangled state and the spin tomogram of the state is found in explicit form. The qubit portrait method is used the Bell inequality violation which provides another tool to prove the property of entanglement of the four-qubit state under consideration.
\end{abstract}

\medskip

\noindent{\bf Keywords:}
qubits, qudits, probability representation of quantum mechanics, quantum tomography, quantum entanglement, bound entanglement, qubit portrait, Smolin state.

\section{Introduction}

\pst
The difference of quantum and classical states is demonstrated by the existance for quantum systems of the phenomenon of entanglement~\cite{Schred}. This phenomenon is closely related to the Einstein-Podolsky-Rosen paradox~\cite{EPR}. The definition of entanglement could be formulated for composite quantum systems as a property of density operator, for example, for bepartite system, determining the system state~\cite{Landay-1927, von-Neuman-1932}. This property distinguishes the presence of classical correlations of the subsystems from strong quantum correlations of the subsystems. For states of two qubits the difference of the correlations could be detected by violation of Bell inequalities~\cite{7,Gisin}.

The simple criterion to detect the entanglement of the state of general system is not known. There exsits a particular criterion~\cite{Peres,Hor} based on partial positive transpose (PPT) of density operator which yields the necessary condition of state separability. The separable states are usually interpreted as the states with classical correlations of the subsystems. The entangled states are usually interpreted as the states with quantum correlations. The PPT-criterion gives the necessary and sufficient criterion of entanglement only for two qubits and for bipartite system of qubit-qutrit.

There exist the states called bound entangled ststes~\cite{Hor-Hor-Hor} with specific behaviour in respect to partial positive transpose of the density matrix. If one can obtain two qubit entangled state from the mixture of qubits applying to this mixture only operations from the set of so-called local operations and classical communications (LOCC)~\cite{Plenio-Virmani} then this state is called distillable. The procedure of creation of entangled pair of qubits out of the mixture is called distillation. Bound entangled states are proved to be not distillable although definitely entangled. Till now the study of bound entangled states is done by using the concrete examples of the composite systems. Our aim is to study the property of bound entaglement. In this paper we will consider Smolin state, the 4-qubit bound entangled state~\cite{8}.

The probability representation of quantum mechanics provides the complete description of quantum system states by means of probability distributions instead of such tradistional ways as wave functions and density matrices~\cite{1,2,3,4}. In particular, spin states can be described by spin tomogram~\cite{3,4}, which is probability for spin projection on a given direction to take its allowed values. This definition can be generalised to describe composite systems comprised of many spins. One of the applications of spin tomogram is studying of quantum entanglement in such systems. The tomographic approach to the problem was presented and applied to various examples in~\cite{5,6}. It allows to test the separability of complex spin states by direct use of the Bell inequalitites, applied to so-called qubit portraits of states. The qubit portrait of Smolin state allows to show that it is not separable across the cut 1 qubit-3 qubits.

The paper is organized as follows. The Section 2 provides the definition of spin tomogram. In the Section 3 the concept of qubit portait of probability distribution is introduced. Then we give the neccessary condition of separability which uses qubit portraits of the distributions associated with spin tomograms of state under examination. In the Section 3 we apply the approach to the case of the Smolin state.

\section{\label{sec:tomogram}Spin tomogram. Tomographic definition of separability.}

\pst By means of spin tomogram any spin state (density matrix) can be invertibly mapped into a set of positive functions. Let us introduce qubit tomogram as two diagonal elements of density matrix in representation parameterized by rotated axis $\vec n$ of quantization
\begin{equation}
\label{tomint:1}
\omega(m,\vec n) = \langle m |\widehat u^{\dag} \widehat{\rho} \widehat u | m \rangle,
\end{equation}
where the unitary matrix $\widehat u$ realises this rotation. The axis and the matrix elements in standard representation can be connected with corresponding Euler angles $\theta$,$\psi$ and $\phi$:
\begin{eqnarray}
\label{tomint:2}
\vec n=(\sin\theta\cos\phi,\sin\theta\sin\phi,\cos\theta),
\nonumber\\
\widehat u =
\left(\begin{array}{ccc}
\cos \frac{\theta}{2}\exp(\frac{i(\phi+\psi)}{2}) & \sin \frac{\theta}{2}\exp(\frac{i(\phi-\psi)}{2})\\
-\sin \frac{\theta}{2}\exp(\frac{i(\psi-\phi)}{2}) & \cos \frac{\theta}{2}\exp(\frac{-i(\phi+\psi)}{2})
\end{array} \right).
\end{eqnarray}
The tomogram is quantum probability of spin having spin projection measurment outcomes, so an associated probability vector can be composed from it, its column containing the probabilities of all possible outcomes:
\begin{equation}
\label{tomint:3}
\vec{\omega}=
\left(\begin{array}{ccc}
\omega(+1/2,\vec n)\\\omega(-1/2,\vec n)
\end{array} \right).
\end{equation}

In the same manner spin tomogram can be defined for qudits of higher dimensions, with spin quantum number $j$. The same formula (\ref{tomint:1}) remains the correct definition, this time $m$ assumed to take on the numbers $-j,...,j$ and the matrices $\widehat u$ belong to irreducible representation of rotation group associated with spin $j$. This tomogram also contains the complete information about state of respective qudit and thus mathematically equivalent to $(2j+1)\times(2j+1)$ density matrix.

Further generalization of tomogram, especially interesting in relation to separability and entanglement problems, can be achieved for systems composed of more than one qudit. In this case a joint probability distribution of all spins projections should be used. For example, for two qudits tomogram 
\begin{equation}
\label{tomint:4}
\omega(m_1,m_2,\vec n_1,\vec n_2) = \langle m_1m_2 | (\widehat u_1\otimes \widehat u_2)^{\dag} \cdot\widehat{\rho} \cdot\widehat u_1\otimes \widehat u_2 | m_1m_2 \rangle
\end{equation}
is a unique marker of state as well. The corresponding probablity vector has a dimension 4:
\begin{equation}
\label{tomint:5}
\vec{\omega}=
\left(\begin{array}{ccc}
\omega(+1/2,+1/2,\vec n_1,\vec n_2)\\\omega(+1/2,-1/2,\vec n_1,\vec n_2)\\\omega(-1/2,+1/2,\vec n_1,\vec n_2)\\\omega(-1/2,-1/2,\vec n_1,\vec n_2)
\end{array} \right).
\end{equation}

The invertibility of state-to-tomogram mapping allows one to rewrite the definition of state separability in purely tomographic terms: two qudits state $\omega(m_1,m_2,\vec n_1,\vec n_2)$ is separable across the cut between its qudits if and only if it can be resolved in a convex sum of products of corresponding one qudit tomograms, i.e.
\begin{equation}
\label{tomint:6}
\omega(m_1,m_2,\vec n_1,\vec n_2) = \sum_{i}p^{(i)}\omega_1^{(i)}(m_1,\vec n_1)\omega_2^{(i)}(m_2,\vec n_2),p^{(i)}\geqslant 0,\sum_{i}p^{(i)}=1.
\end{equation}
The general statement holds, that separability across the cut between subsytems $A$ and $B$ can be defined as possibility to decompose its tomogram in a convex sum of respective products of tomograms of $A$ and $B$.

\section{\label{sec:portrait}Qubit portrait and necessary condition for separability.}

\pst
The joint tomographic distribution of two qubits (\ref{tomint:5}) can be used directly to test if given state is separable via finding if it satisfies the Bell-CHSH inequality (it also commonly refered to as Bell inequality)~\cite{7}. The inequality holds that for correlation coefficients between parameters $a$ and $b$, both taking on numbers $1$ and $-1$,(denoted $C(a,b)$) in classical probability should be true that
\begin{equation}
\label{tomint:6.5}
B=|C(a1,b1)+C(a1,b2)+C(a2,b1)-C(a2,b2)|\leqslant 2
\end{equation}
where the values $a1$ and $b1$ each can be equal $1$ or $-1$ independently. Particularly the inequality is satisfied in case of two qubit system in separable state, doubled spin projections of two spins on any directions being chosen as the parameters $a$ and $b$.

Thus one can use compliance of a state with the inequality as a necessary criterion of its separability. The correlations coefficients which are dealt with in the inequality can be expressed in terms of tomogram. Namely, if one introduces a stochastic matrix
\begin{equation}
\label{tomint:7}
M(\vec a,\vec b,\vec c,\vec d)=
\left(\begin{array}{cccc}
\omega(+1/2,+1/2,\vec a,\vec b) & \omega(+1/2,+1/2,\vec a,\vec c) & \omega(+1/2,+1/2,\vec d,\vec b) & \omega(+1/2,+1/2,\vec d,\vec c)\\
\omega(+1/2,-1/2,\vec a,\vec b) & \omega(+1/2,-1/2,\vec a,\vec c) & \omega(+1/2,-1/2,\vec d,\vec b) & \omega(+1/2,-1/2,\vec d,\vec c)\\
\omega(-1/2,+1/2,\vec a,\vec b) & \omega(-1/2,+1/2,\vec a,\vec c) & \omega(-1/2,+1/2,\vec d,\vec b) & \omega(-1/2,+1/2,\vec d,\vec c)\\
\omega(-1/2,-1/2,\vec a,\vec b) & \omega(-1/2,-1/2,\vec a,\vec c) & \omega(-1/2,-1/2,\vec d,\vec b) & \omega(-1/2,-1/2,\vec d,\vec c)
\end{array} \right)
\end{equation}
then for any separable two qubit state $\omega(m_1, m_2,\vec n_1,\vec n_2)$ and for any unit vectors $\vec a$,$\vec b$,$\vec c$ and $\vec d$ it shall be true that
\begin{eqnarray}
\label{tomint:8}
B=|M_{11}-M_{21}-M_{31}+M_{41}+M_{12}-M_{22}-M_{32}+M_{42}+
\nonumber\\
+M_{13}-M_{23}-M_{33}+M_{43}-M_{14}+M_{24}+M_{34}-M_{44}|\leqslant 2.
\end{eqnarray}
If there is a combination of $\vec a$,$\vec b$,$\vec c$ and $\vec d$ which breakes the inequality than the state is not separable.

The authors of ~\cite{5} developed an ansatz which allows similarly to test more general qudit states of higher dimensions for separability which is also based on the Bell iequality. To build it one needs to introduce a concept of qubit portraits of probability distributions.

Let us consider a discrete probability distribution $(p_1,...,p_n)$. By means of linear transform it can be reduced to 2-dimensional distribution:
\begin{equation}
\label{tomint:9}
q_i=\sum_{j=1,n}\pi_{ij}p_j,i=1,2.
\end{equation}
If matrix $\pi$ satisfies the conditions that $\sum_{i=1,2}\pi_{ij}=1$ and $\pi_{ij}\geqslant 0$ then every distribution is transformed to correct normalized and positive distribution. Any such linear compression is called \emph{qubit portrait} of a given distribution. One can make a qubit portraits of two tomograms $\omega_A(m_A,...)$ and $\omega_B(m_B,...)$ associated with systems A and B:
\begin{eqnarray}
\label{tomint:10}
w_{A}(i,...)=\sum_{m_A}\pi_{im_A}^{(A)}\omega_A(m_A,...),
\nonumber\\
w_{B}(i,...)=\sum_{m_B}\pi_{im_B}^{(B)}\omega_B(m_B,...),
\nonumber\\
i=1,2.
\end{eqnarray}

Let us have a joint tomogram for a system unifying A and B $\omega(m_A,m_B,...)$ (see the formula (\ref{tomint:4})) which is separable according to the difinition (\ref{tomint:6}). The observation can be made that if one applies the direct product of the operations (\ref{tomint:10}) to joint tomographic distribution for systems A and B the property of separability remains true:
\begin{equation}
\label{tomint:11}
w(i,j,...)=\sum_{m_A}\sum_{m_B}(\pi_{im_A}^{(A)}\cdot \pi_{jm_B}^{(B)})\omega(m_A,m_B,...)=\sum_{k}p^{(k)}\left(\sum_{m_A}\pi_{im_A}^{(A)}\omega_A^{(k)}(m_A,...)\right)\left(\sum_{m_B}\pi_{im_B}^{(B)}\omega_B^{(k)}(m_B,...)\right).
\end{equation}
As a result we have a family of joint probability distributions for pair of indeces $\{i=0,1,j=1,2\}$ which is separable:
\begin{equation}
\label{tomint:12}
w(i,j,...)=\sum_{k}p^{(k)}w_{A}^{(k)}(i,...)w_{B}^{(k)}(j,...),
\end{equation}
hence it is still subject to the Bell inequality constraint:
\begin{eqnarray}
\label{tomint:13}
B=|w(0,0,...)-w(0,1,...)-w(1,0,...)+w(1,1,...)
+w(0,0,...)-w(0,1,...)-w(1,0,...)+w(1,1,...)+
\nonumber\\
+w(0,0,...)-w(0,1,...)-w(1,0,...)+w(1,1,...)
-w(0,0,...)+w(0,1,...)+w(1,0,...)-w(1,1,...)|\leqslant 2.
\end{eqnarray}

We can state the neccessary condition of separability: if a joint state of systems A and B is separable then for any qubit portrait (\ref{tomint:10}) and any set of parameters the Bell inequality is true.

\section{\label{sec:smolin}Smolin state.}

\pst Smolin introduced ~\cite{8} a state of four qubits A,B,C and D with interesting properties from the point of quantum information:
\begin{eqnarray}
\label{smolin:1}
\widehat{\rho}_S=\frac{1}{4}(
|\Phi_{AB}^+\rangle \langle\Phi_{AB}^+|\otimes|\Phi_{CD}^+\rangle \langle\Phi_{CD}^+|+
|\Phi_{AB}^-\rangle \langle\Phi_{AB}^-|\otimes|\Phi_{CD}^-\rangle \langle\Phi_{CD}^-|+
\nonumber\\
|\Psi_{AB}^+\rangle \langle\Psi_{AB}^+|\otimes|\Psi_{CD}^+\rangle \langle\Psi_{CD}^+|+
|\Psi_{AB}^-\rangle \langle\Psi_{AB}^-|\otimes|\Psi_{CD}^-\rangle \langle\Psi_{CD}^-|
).
\end{eqnarray}
The notation used in this formula:
\begin{eqnarray}
\label{smolin:2}
|\Phi_{AB}^{\pm}\rangle=\frac{1}{\sqrt{2}}(|m_A=+1/2,m_B=+1/2\rangle\pm|m_A=-1/2,m_B=-1/2\rangle),
\nonumber\\
|\Psi_{AB}^{\pm}\rangle=\frac{1}{\sqrt{2}}(|m_A=+1/2,m_B=-1/2\rangle\pm|m_A=-1/2,m_B=+1/2\rangle).
\end{eqnarray}
$|\Phi_{CD}^{\pm}\rangle$ and $\Psi_{CD}^{\pm}\rangle$ refer to the same 2-qubit statets but of qubits C and D. The form in which this state is introduced suggests that it is separable across the cut AB:CD. Further analysis shows that in fact Smolin state is symmetric under permutation of any pair of qubits. In standard qubit basis the state has a form
\begin{equation}
\label{smolin:3}
\widehat{\rho}_S=\frac{1}{16}\left(\widehat{I}+\sum_{i=x,y,z}\widehat{\sigma_i}^{(A)}\otimes\widehat{\sigma_i}^{(B)}\otimes\widehat{\sigma_i}^{(C)}\otimes\widehat{\sigma_i}^{(D)}\right).
\end{equation}
Smolin state has an intersesting properties in respect to quantum information theory. Being the mixture of different two-qubit entangled states, it however can't be distilled to one of this states without at least two of the qubits placed in one laboratory. Such states are called bound entangled. We will prove that the state is entangled across the cat 1 qubit:3 qubits using the method described in the previous section.

The 4-qubit tomogram of Smolin state is
\begin{equation}
\label{smolin:4}
\omega(m_A,m_B,m_C,m_D,\vec n_A,\vec n_B,\vec n_C,\vec n_D)=\frac{1}{16}+m_Am_Bm_Cm_D\sum_{i=x,y,z}n_i^{(A)}n_i^{(B)}n_i^{(C)}n_i^{(D)}.
\end{equation}
We consider separation across the cut A:BCD. The tomogram of the first subsystem is already a qubit portrait of itself. Three qubits subsystem will undergo the following transformation to qubit portrait: the sum of tomogram components for which $m_Bm_Cm_D=1/8$ is to be the first component of the portrait, sum of those for which $m_Bm_Cm_D=-1/8$ - the second. Applying this compression to the tomogram leads to the following  portrait:
\begin{equation}
\label{smolin:4}
\omega(m_A,p,\vec n_A,\vec n_B,\vec n_C,\vec n_D)=\frac{1}{4}+m_Ap\frac{1}{2}\sum_{i=x,y,z}n_i^{(A)}n_i^{(B)}n_i^{(C)}n_i^{(D)}
\end{equation}
where $p$ gets the number 1 (for the sum of elements with $m_Bm_Cm_D=1/8$) or -1 (for the sum of elements with $m_Bm_Cm_D=-1/8$). One can see that the parameters $p$ and $m_A$ replace the parameters $i$ and $j$ from the definition (\ref{tomint:12}) as quantum numbers for ``qubits''. The elements of stochastic matrix $M$ for this distribution reads
\begin{eqnarray}
\label{smolin:5}
M_{11}=\frac{1}{4}+\frac{1}{4}\sum_{i=x,y,z}a_i^{(A)}b_i^{(B)}b_i^{(C)}b_i^{(D)},
\nonumber\\
M_{21}=\frac{1}{4}-\frac{1}{4}\sum_{i=x,y,z}a_i^{(A)}b_i^{(B)}b_i^{(C)}b_i^{(D)},
\nonumber\\
M_{31}=\frac{1}{4}-\frac{1}{4}\sum_{i=x,y,z}a_i^{(A)}b_i^{(B)}b_i^{(C)}b_i^{(D)},
\nonumber\\
M_{41}=\frac{1}{4}+\frac{1}{4}\sum_{i=x,y,z}a_i^{(A)}b_i^{(B)}b_i^{(C)}b_i^{(D)},
\nonumber\\
M_{12}=\frac{1}{4}+\frac{1}{4}\sum_{i=x,y,z}a_i^{(A)}c_i^{(B)}c_i^{(C)}c_i^{(D)},
\nonumber\\
M_{22}=\frac{1}{4}-\frac{1}{4}\sum_{i=x,y,z}a_i^{(A)}c_i^{(B)}c_i^{(C)}c_i^{(D)},
\nonumber\\
M_{32}=\frac{1}{4}-\frac{1}{4}\sum_{i=x,y,z}a_i^{(A)}c_i^{(B)}c_i^{(C)}c_i^{(D)},
\nonumber\\
M_{42}=\frac{1}{4}+\frac{1}{4}\sum_{i=x,y,z}a_i^{(A)}c_i^{(B)}c_i^{(C)}c_i^{(D)},
\nonumber\\
M_{13}=\frac{1}{4}+\frac{1}{4}\sum_{i=x,y,z}d_i^{(A)}b_i^{(B)}b_i^{(C)}b_i^{(D)},
\nonumber\\
M_{23}=\frac{1}{4}-\frac{1}{4}\sum_{i=x,y,z}d_i^{(A)}b_i^{(B)}b_i^{(C)}b_i^{(D)},
\nonumber\\
M_{33}=\frac{1}{4}-\frac{1}{4}\sum_{i=x,y,z}d_i^{(A)}b_i^{(B)}b_i^{(C)}b_i^{(D)},
\nonumber\\
M_{43}=\frac{1}{4}+\frac{1}{4}\sum_{i=x,y,z}d_i^{(A)}b_i^{(B)}b_i^{(C)}b_i^{(D)},
\nonumber\\
M_{14}=\frac{1}{4}+\frac{1}{4}\sum_{i=x,y,z}d_i^{(A)}c_i^{(B)}c_i^{(C)}c_i^{(D)},
\nonumber\\
M_{24}=\frac{1}{4}-\frac{1}{4}\sum_{i=x,y,z}d_i^{(A)}c_i^{(B)}c_i^{(C)}c_i^{(D)},
\nonumber\\
M_{34}=\frac{1}{4}-\frac{1}{4}\sum_{i=x,y,z}d_i^{(A)}c_i^{(B)}c_i^{(C)}c_i^{(D)},
\nonumber\\
M_{44}=\frac{1}{4}+\frac{1}{4}\sum_{i=x,y,z}d_i^{(A)}c_i^{(B)}c_i^{(C)}c_i^{(D)},
\nonumber\\
\end{eqnarray}
The Bell number is
\begin{equation}
\label{smolin:6}
B=\left|\sum_{i=x,y,z}\left((a_i^{(A)}+d_i^{(A)})\prod_{k=B,C,D}b_i^{(k)}+(a_i^{(A)}-d_i^{(A)})\prod_{k=B,C,D}c_i^{(k)}\right)\right|.
\end{equation}
In this formula we used the notation $b_i^{(k)}$ and $c_i^{(k)}$ for the numbers which the unit vector parameters $n_i^{(B)}$, $n_i^{(C)}$ and $n_i^{(D)}$ take on. It is the core difference between the parametrization of usual two qubit system and our qubit portrait that in the first case one unit vector is associated with one qubit while in the second case there are three vectors each verying independently. Hence our goal is to provide the values for 8 unit vectors $\vec a$, $\vec b^{(B)}$, $\vec b^{(C)}$, $\vec b^{(D)}$, $\vec d$, $\vec c^{(B)}$, $\vec c^{(C)}$, $\vec c^{(D)}$. One can see that, for example, for the combination $\vec a^{(A)}=(1/\sqrt{2},-1/\sqrt{2},0)$,$\vec d^{(A)}=(1/\sqrt{2},1/\sqrt{2},0)$,$\vec b^{(k)}=(1,0,0)$,$\vec c^{(k)}=(0,1,0)$,$k=1,2,3$ $B$ reaches $2\sqrt{2}$ hence Smolin state has entanglement between qubit and the subsystem of three other qubits.

\section{Summary}

\pst
To conclude we resume the main results of the paper. We constructed the spin tomogram of the four qubit Smolin state which is known as example of bound entangled state. The entanglement of this state was proved by a specific method which is qubit portrait method providing a tool to study correlations in composite systems of qudits reducing the the spin-tomographic probabilities of the initial system state to probability vector of two qubit state. We have shown that the constructed qubit portrait of Smolin state violates Bell inequality which means that the state is entangled.


\end{document}